\documentclass{emulateapj}

\usepackage{array}
\usepackage{color}

\begin{document}

\newcommand{\kms}{km~s$^{-1}$}
\newcommand{\ms}{~M$_{\sun}$}
\newcommand{\egs}{erg~egs$^{-1}$}

\newcommand{\equ}[1]{eq.~(\ref{eq:#1})}
\newcommand{\equs}[1]{eqs.~(\ref{eq:#1})}
\newcommand{\equm}[1]{(\ref{eq:#1})}
\newcommand{\Equ}[1]{Eq.~(\ref{eq:#1})}
\newcommand{\Equs}[1]{Eqs.~(\ref{eq:#1})}
\newcommand{\equnp}[1]{eq.~\ref{eq:#1}}
\newcommand{\se}[1]{\S\ref{sec:#1}}
\newcommand{\fig}[1]{Fig.~\ref{fig:#1}}
\newcommand{\figs}[1]{Figs.~\ref{fig:#1}}
\newcommand{\Fig}[1]{Figure~\ref{fig:#1}}
\newcommand{\Figs}[1]{Figures~\ref{fig:#1}}
\newcommand{\be}{\begin{equation}}
\newcommand{\ee}{\end{equation}}
\newcommand{\bea}{\begin{eqnarray}}
\newcommand{\eea}{\end{eqnarray}}
\def\ra{\rangle}
\def\la{\langle}

\newcommand{\no}{\noindent}
\newcommand{\bk}{\hfill\break}
\newcommand{\msun}{\mathrm{M}_\odot}
\newcommand{\msolar}{\mathrm{M}_{\odot}}
\newcommand{\lsun}{L_\odot}
\newcommand{\ifm}[1]{\relax\ifmmode#1\else$\mathsurround=0pt #1$\fi}
\newcommand{\hmpc}{\,\ifm{h^{-1}}{\rm Mpc}}
\newcommand{\hkpc}{\,\ifm{h^{-1}}{\rm kpc}}
\newcommand{\Mpc}{\,{\rm Mpc}}
\newcommand{\kpc}{\,{\rm kpc}}
\newcommand{\pc}{\,{\rm pc}}
\newcommand{\Gyr}{\,{\rm Gyr}}
\newcommand{\gyr}{\,{\rm Gyr}}
\newcommand{\Myr}{\,{\rm Myr}}
\newcommand{\yr}{\,{\rm yr}}
\newcommand{\ergs}{\,{\rm erg}\,{\rm s}^{-1}}

\newcommand{\ltsima}{$\; \buildrel < \over \sim \;$}
\newcommand{\lsim}{\lower.5ex\hbox{\ltsima}}
\newcommand{\gtsima}{$\; \buildrel > \over \sim \;$}
\newcommand{\gsim}{\lower.5ex\hbox{\gtsima}}
\newcommand{\prop}{\propto}
\newcommand{\dd}{\rm d}
\newcommand{\pa}{\partial}
\newcommand{\const}{\rm const.}
\newcommand{\rar}{\rightarrow}
\newcommand{\lar}{\leftarrow}

\def\secpush{\vskip0pt plus 2.0\baselineskip\penalty-250
             \vskip0pt plus -2.0\baselineskip }

\def\sy{\mathrm{M}_\odot\, {\rm yr}^{-1}}
\def\Mpcc{\,{\rm Mpc}^{-3}}
\def\cms{\,{\rm cm}^{-2}}
\def\cmc{\,{\rm cm}^{-3}}

\def\col#1{\colhead{#1}}

\def\omm{\Omega_{\rm m}}
\def\oml{\Omega_{\Lambda}}

\def\Mv{M_{\rm v}}
\def\Mvd{\dot{M}_{\rm v}}
\def\Rv{R_{\rm v}}
\def\Vv{V_{\rm v}}
\def\Tv{T_{\rm v}}
\def\Mg{M_{\rm g}}

\def\Mt{M_{\rm tot}}
\def\Mb{M_{\rm b}}
\def\Md{M_{\rm d}}
\def\Rd{R_{\rm d}}
\def\fg{f_{\rm g}}

\def\Mc{M_{\rm c}}
\def\Rc{R_{\rm c}}
\def\Rci{R_{\rm c,i}}
\def\Vc{V_{\rm c}}
\def\Vrot{V_{\rm rot}}
\def\Vcc{V_{\rm c,circ}}
\def\Vci{V_{\rm c,i}}
\def\Sigmac{\Sigma_{\rm c}}
\def\sigmac{\sigma_{\rm c}}
\def\rt{\tilde{r}}

\def\td{t_{\rm d}}
\def\fdis{f_{\rm dis}}
\def\tdis{t_{\rm dis}}

\newcommand{\adg}[1]{\textcolor{green}{#1}}
\newcommand{\adb}[1]{\textcolor{blue}{#1}}
\newcommand{\adr}[1]{\textcolor{red}{[AD: #1]}}
\newcommand{\fbr}[1]{\textcolor{magenta}{[FB: #1]}}


\title{Black Hole growth and AGN obscuration by instability-driven inflows
in high-redshift disk galaxies fed by cold streams}

\author{Fr\'ed\'eric Bournaud}
\affil{CEA, IRFU, SAp, 91191 Gif-sur-Yvette, France; frederic.bournaud@cea.fr}

\author{Avishai Dekel}
\affil{Racah Institute of Physics, The Hebrew University, Jerusalem 91904, Israel; dekel@phys.huji.ac.il}

\author{Romain Teyssier}
\affil{CEA, IRFU, SAp, 91191 Gif-sur-Yvette, France.\\
Institute for Theoretical Physics, University of Z\"urich, CH-8057 Z\"urich, Switzerland.}

\author{Marcello Cacciato}
\affil{Racah Institute of Physics, The Hebrew University, Jerusalem 91904, Israel.}

\author{Emanuele Daddi}
\affil{CEA, IRFU, SAp, 91191 Gif-sur-Yvette, France.}

\author{St\'ephanie Juneau}
\affil{Steward Observatory, University of Arizona, Tucson, AZ 85721; sjuneau@as.arizona.edu }

\author{Francesco Shankar}
\affil{Max-Planck-Instit\"{u}t f\"{u}r Astrophysik, Karl-Schwarzschild-Str. 1, D-85748, Garching, Germany.\\}


\begin{abstract}
Disk galaxies at high redshift have been predicted to maintain high gas surface densities due to continuous feeding by intense cold streams leading to violent gravitational instability, transient features and giant clumps. Gravitational torques between the perturbations drive angular momentum out and mass in, and the inflow provides the energy for keeping strong turbulence. 
We use analytic estimates of the inflow for a self-regulated unstable disk at a Toomre stability parameter $Q\sim 1$, and isolated galaxy simulations capable of resolving the nuclear inflow down to the central parsec. We predict an average inflow rate $\sim\!10\,\sy$ through the disk of a $10^{11}\,\msun$ galaxy, with conditions representative of $z\!\sim\!2$ stream-fed disks. The inflow rate scales with disk mass and $(1+z)^{3/2}$. It includes clump migration and inflow of the smoother component, valid even if clumps disrupt. This inflow grows the bulge, while only a fraction $\gsim\!10^{-3}$ of it needs to accrete onto a central black hole (BH), in order to obey the observed BH-bulge relation. A galaxy of $10^{11}\msun$ at $z\!\sim\!2$ is expected to host a BH of $\sim\! 10^8\msun$, accreting on average with moderate sub-Eddington luminosity $L_\mathrm{X}\!\sim\!10^{42-43}\,\ergs$, accompanied by brighter episodes when dense clumps coalesce. We note that in rare massive galaxies at $z\!\sim\!6$, the same process may feed $\sim\!10^9\,\msun$ BH at the Eddington rate. High central gas column densities can severely obscure AGN in high-redshift disks, possibly hindering their detection in deep X-ray surveys.
\end{abstract}


\section{Introduction}
The growth of massive black holes (BH) and the associated Active Galactic Nuclei (AGN) are commonly assumed to be driven by gas inflows from galaxy mergers \citep{DM05,H06}. However, baryons are fed into high-redshift galaxies through cold streams, in which the contribution of smooth gas and small clumps is larger than that of mergers of mass ratio greater than 1:10 \citep{dekel09, brooks09}.

Cold accretion and high gas fractions in high-redshift disks trigger a violent instability, often characterized by giant star-forming clumps. Using analytic estimates (Section~2) and simulations (Section~3), we address the possibility that the gas supply for BH growth is provided by instability-driven inflows, and evaluate the obscuration of AGN in this model (Section~4).


\newpage

\section{Inflow in high-redshift unstable disks}
Many $z\gsim 1$ galaxies of stellar masses $\approx \!10^{10-11}\,\msun$ are rotating disks with giant clumps of $\sim \! 10^{8-9}\msun$ each \citep{E07, genzel11}. In most cases, on-going mergers are not favored by gas kinematics \citep{genzel08} and photometric properties \citep{E09}. Instead, high gas fractions of $\sim 50\%$ \citep{daddi10, tacconi10} and high gas velocity dispersions \citep{FS09} are consistent with gravitational instability and self regulation at $Q \! \sim \! 1$.

\medskip

The occurrence and persistence of gravitational disk instability at high redshift is a natural result of the high surface density of the cold disk component, which is maintained by the continuous gas supply through cold cosmic streams \citep{dekel09}. The instability is predicted to be self regulated in steady state for a few Gyr \citep*[][DSC09]{DSC09}, and cosmological simulations reveal clumpy disks that resemble the observed ones \citep{agertz,ceverino}. If giant clumps survive stellar feedback for a few $10^8$~yr, they migrate into a central bulge \citep{noguchi, BEE07}, as supported by observations \citep{genzel08}.
Typical clumps seem to survive \citep{KD10,genzel11}, while the most luminous clumps could undergo faster disruption \citep{murray10,genzel11}. Nevertheless, the inflow is a robust feature of the disk instability, driven by gravity torques independently of the survivability of bound clumps \citep[][DSC09]{gammie01}.

\medskip

A key element in the self-regulated instability at $Q\! \sim \! 1$, given the
high surface density, is maintaining strong turbulence, with a velocity
dispersion $\sim\!50$\kms for circular velocity $\sim\!200$\kms, as observed in 
high-$z$ disks. Supersonic turbulence dissipates in a few disk vertical
crossing times \citep{mclow99}, and should therefore be continuously powered.
An obvious source is the gravitational energy released when mass flows toward the center (DSC09; \citealt{EB10}). Assuming that the incoming streams deposit mass and energy at the outskirts of
the disk where they do not contribute directly to driving disk turbulence, as
is suggested by cosmological simulations \citep{agertz, ceverino}, the main energy source
is the inflow within the disk \citep{KB10}. 
This energy is transferred to velocity dispersion by gravitational torques involving massive clumps, transient features and smoother regions of the disk. 
Stellar feedback is unlikely to be the main driver of the turbulence, e.g., because the velocity dispersion is observed not to be tightly correlated with proximity to star-forming clumps \citep{FS09,genzel11}.

\medskip

The baryonic inflow rate $\dot{M}_{\rm b}$ is estimated to be:
\be
\dot{M}_{\rm b}\sim 0.2\frac{\Md}{\td}\left(\frac{\sigma}{V}\right)^2 \, ,
\label{eq:Mdot0}
\ee
where $\Md$ is the disk mass, $\td$ is the crossing time, $V$ is the circular
velocity and $\sigma$ is the 1D velocity dispersion. At this rate, the power
released by the inflow down the potential gradient, $\dot{M}_{\rm b}V^2$,
compensates for the dissipative loss rate, $\Mg\sigma^2/\tdis$, where $\Mg\sim
0.5\Md$ is the disk gas mass and $\tdis \sim 3\td$ \citep{Cacciato11} \footnote{we used $0.5/3\!\approx\!0.2$ in Eq.~1}. 
Similar estimates are obtained from the rate of energy exchange by
clump encounters (DSC09, eqs.~21 and 7), and from the angular-momentum exchange
among the transient perturbations in a viscous disk \citep[][DSC09, eq.
24]{gammie01,genzel08}.
Adopting $\sigma/V\! \sim \! 0.2$ and 
$\td \!\sim \!30$~Myr at $z=2$, the inflow rate is 
\be
\dot{M}_{\rm b} \simeq 25 M_{\rm b,11} (1+z)_3^{3/2}  \, ,
\label{eq:Mdot}
\ee
where $M_{\rm b,11}$ is the disk mass in $10^{11}\,\msun$, and $(1+z)_3=(1+z)/3$. Note that the $(1+z)_3^{3/2}$ scaling corresponds to the disk crossing time, proportional to the halo crossing time at fixed overdensity in a matter-dominated Universe.

\medskip 

A continuity equation for the disk gas, being drained into star formation, 
outflows and bulge and replenished by streams, yields at $z \leq 3$ 
a steady state with $\dot{M}_{\rm b}$ about one third of 
the cosmological baryonic accretion rate \citep{bouche10,KD11}. With the 
latter being $\sim\!80\,M_{12}\,(1+z)_3^{5/2} \sy$ \citep[][$M_{12}$ being the halo mass in $10^{12}\,\msun$]{neistein06}, this estimate is consistent with \Equ{Mdot} at $z=3$, with a different redshift dependence for the {\em cosmological} inflow rate.
 
\section{Simulations}

\begin{table*}
\centering
\caption{Simulation parameters, average gas inflow rates $\dot{M}_{\rm g}$ between $t=250$ and 400~Myr, and predicted value $f_\mathrm{g} \dot{M}_{\rm b}$ from Equation~\ref{eq:Mdot}, in $\sy$.}\label{tab1}
\begin{tabular}{lcccccccccc}
\hline
\hline
ID   &$\;$& $\epsilon_\mathrm{AMR}$ & $M_*$($\times$10$^{10}$\ms) & $R_\mathrm{d}$ (kpc) & $f_\mathrm{bulge}$ & $\eta_{SN}$ &$ \;\;$ & $\dot{M}_{\rm g}$(150~pc) & $\dot{M}_{\rm g}$(1~kpc) & $f_\mathrm{g} \dot{M}_{\rm b}$ \\
\hline
HR1 && 1.7~pc    & $7.0$ & 16 &  14\%  &0.5    && 5.6  & 7.2    &  15.0  \\
MR1 && 10.2~pc  & $7.0$ & 16 &  14\%  &0.5    && 5.2 &  6.8  &  15.0  \\
HR2 && 3.4~pc    & $1.5$ &  7 &   8\%  &0.5      && 1.2  & 2.9    &   3.2   \\
MR2 && 10.2~pc  & $1.5$ &  7 &   8\%  &0.5      && 1.3  & 3.1    &   3.2  \\
MR3 && 10.2~pc  & $3.3$ & 12 &   0\%  &0.5      && 3.4  & 6.5   &   8.2   \\
MR4 && 10.2~pc  & $7.0$ & 16 &  14\% & 0.5-5 && 4.4  & 6.9   & 15.0   \\
\hline
\end{tabular}
\end{table*}

\begin{figure*}
\centering
\vspace{1cm}
\includegraphics[width=15cm]{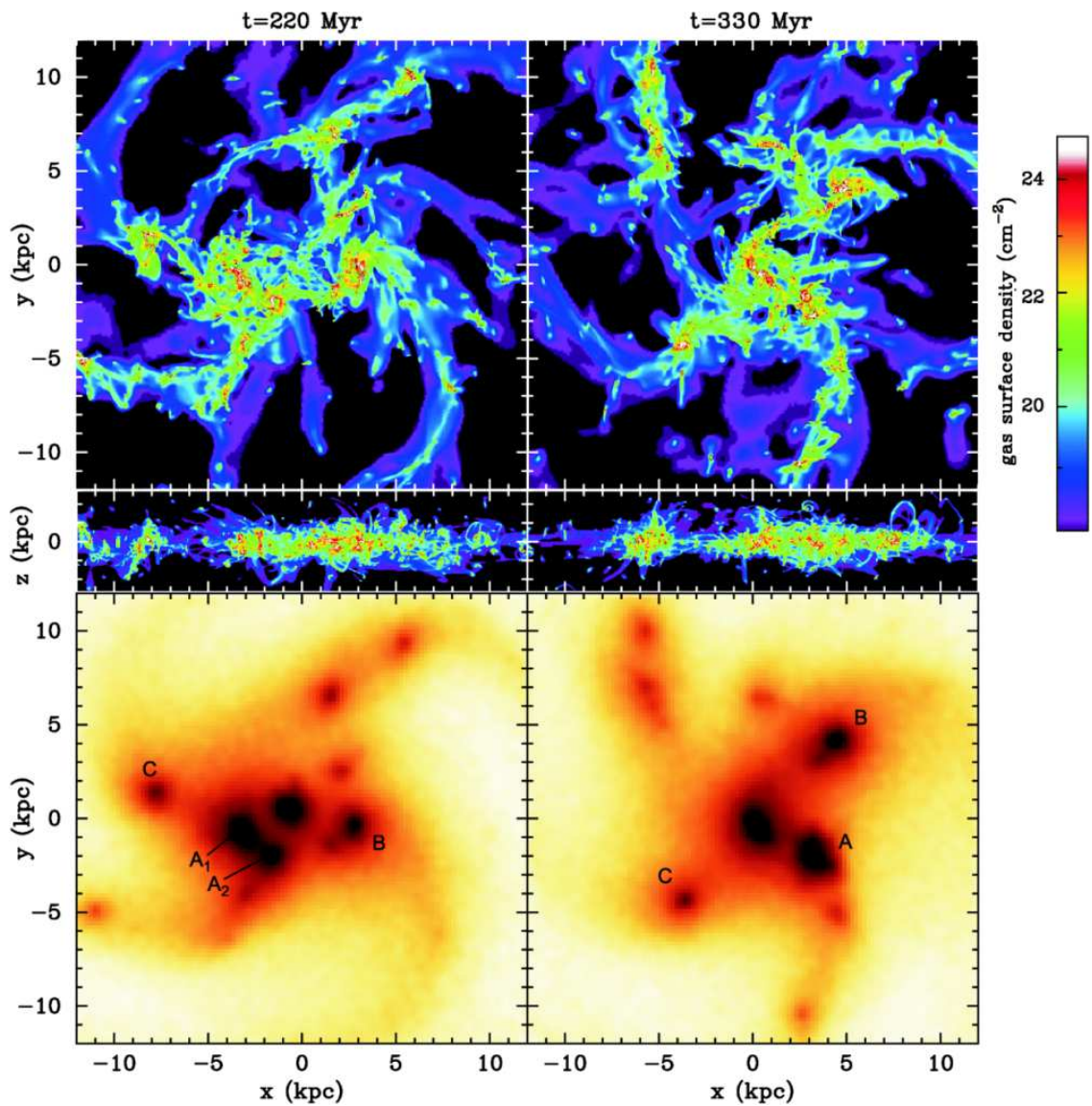}
\caption{Surface density of gas and stars at two snapshots of simulation HR1. The gas is shown face-on (top) and edge-on (middle), the stars are shown face-on (bottom). To mimic optical images, the luminosity of each stellar particle is constant during its first 10~Myr then decreases as $t^{-0.7}$. Clumps A$_\mathrm 1$ and A$_\mathrm 2$ merge together between the snapshots.}
\label{fig1}
\end{figure*}

We perform six idealized simulations using the Adaptive Mesh Refinement (AMR) code
RAMSES \citep{teyssier02}, with a density-dependent grid refinement
strategy and a barotropic cooling model producing realistic phase-space ISM structure \citep{bournaud10}. The initial parameters of the six simulations are listed in Table~\ref{tab1}. The smallest cell size is $\epsilon_\mathrm{AMR}$=1.7 or
3.4~pc in high-resolution (HR) runs and 10.2~pc in medium-resolution (MR) runs.
Cells are refined if they contain gas mass larger than $2 \times
10^3$~M$_\sun$, or more than 20 particles, or if the Jeans length is resolved by less than four
cells. The initial number of dark-matter and stellar particles are $4 \times 10^6$ and $6 \times 10^5$ in HR
and MR, respectively. Star formation occurs above a density threshold of $5 \times 10^4$ and $4 \times 10^3$~cm$^{-3}$ for HR and MR, with a star-formation efficiency per local free-fall time of 4\%. A fraction of the supernovae energy
$\eta_\mathrm{SN}=0.5$ is released as kinetic feedback. For MR4, feedback is boosted up after 250~Myr to $\eta_\mathrm{SN}=5$ in order to disrupt the clumps.

\medskip

The simulations start as pre-formed disks in dark matter halos.
The exponential disk is truncated at $R_\mathrm{d}$, with scale-length
$R_\mathrm{d}/4$ and $R_\mathrm{d}/1.5$ for stars and gas, respectively. The
dark-matter halo has a Burkert profile with a scale-length
$R_\mathrm{d}/3$. The dark-matter mass fraction within $R_\mathrm{d}$ is 0.25.
The key parameter distinguishing high-redshift disks is the gas fraction, set to
50\%, as observed at $z\sim2$. A small bulge containing a fraction
$f_\mathrm{bulge}$ of the stellar mass is assumed.

\medskip

Within one orbital time, the violent gravitational instability, with $Q\!<\!1$ in the initial conditions, develops giant baryonic clumps and supersonic turbulence. The simulations resolve dense gas structures up to $\gsim 10^7$~cm$^{-3}$, with a turbulence cascade in the ISM probed by the spectral analysis of model HR1 \citep[][Section~3.4 and Fig.~15]{bournaud10}. These isolated models allow us to follow the evolution of unstable disks for a couple of orbital times, i.e. a few $10^8$~yr, after which external gas supply should be included.

\begin{figure}
\centering
\includegraphics[width=5cm,angle=270]{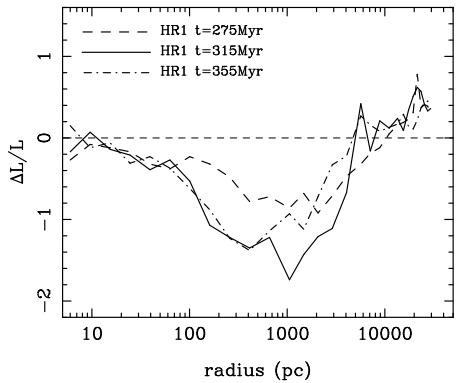}
\caption{Relative change in angular momentum per rotation period at radius $r$
in the disk of run HR1, as evaluated from the gravitational torques.}
\label{fig-torques}
\end{figure}

\bigskip

These simulations demonstrate that the instability-driven inflow is a generic
feature, not limited to the migration of clumps. Figure~\ref{fig1} shows two
snapshots of run HR1. At $t=210$~Myr, giant clumps have formed
in the outer disk and the turbulent speed has reached a steady state with a
mass-weighted average $\sigma \simeq 65$~km~s$^{-1}$. About half of the gas mass
is located between the clumps, the central kpc containing initially only
$3\times 10^7$~M$_{\sun}$ of gas. At $t=330$~Myr,
the giant clumps have barely started their inward migration but the
central kpc already contains $1.5\times10^9\msun$ of originally diffuse gas that
was driven in by torques.

Figure~\ref{fig-torques} shows the relative change in angular momentum per
rotation period due to the gravity torques, $\Delta L/L = 2 \pi F_t/F_r$, where
$F_t$ and $F_r$ are the tangential and radial components of the force, averaged
in radial bins. In the inner 4~kpc, most angular momentum is removed within one
rotation, demonstrating the efficiency torquing driving the inflow. 

\medskip

Table~1 lists the gas inflow rates through spherical shells at $r\!=\!1$~kpc 
and 150~pc, averaged between $t\!=\!250$ and 450~Myr, i.e. between the onset of 
clump formation and their central coalescence. The simulated gas inflow rate at 1~kpc
is in the ballpark of the theoretical estimate of Eq.~\ref{eq:Mdot}, smaller by
up to a factor of two, probably because the replenishment by fresh gas is
ignored in the simulations. The inflow rate is not affected by the stronger
feedback and clump disruption in run MR~4. Indeed, in the simulation by
\citet{genel} in which the clumps are disrupted by enhanced outflows, the inflow
rate is similar to that in their reference simulation where the clumps survive,
reaching in both cases a peak gas surface density higher than $10^3$\ms~pc$^{-2}
\approx 10^{23}{\rm cm}^{-2}$.

\medskip

\begin{figure}
\centering
\includegraphics[width=7cm]{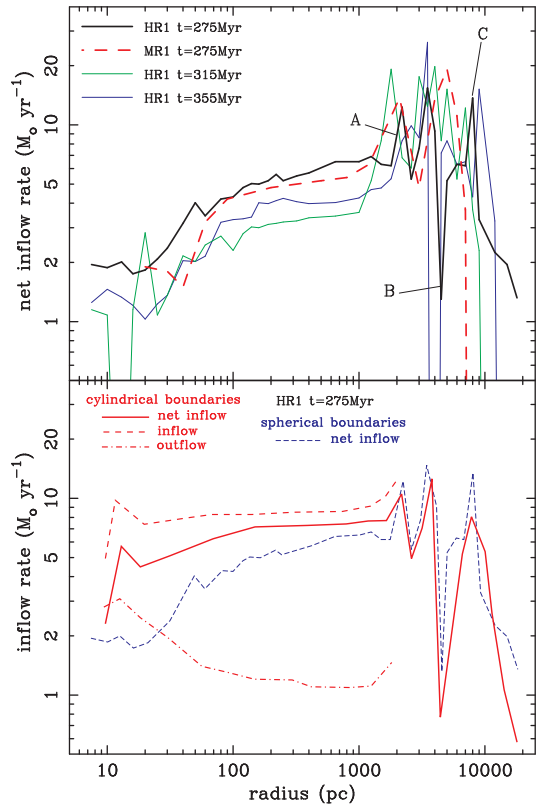}
\caption{Radial profiles of the gas inflow rate. {\bf Top: } net inflow (inflow minus outflow) through spherical boundaries for run HR1 at three different times and for run MR1. It is dominated by clump migration at large radii (note that clump B is moving outward), and by smooth inflow of $1-7\sy$ between 70~pc and 2~kpc. Below $\sim 30-50$~pc the inflow rate drops, but it remains at $1-2\sy$ in the inner 30~pc. {\bf Bottom: } Net inflow rate through cylindrical and spherical boundaries for HR1, comparing inflow and outflow. The drop in the net spherical inflow rate at small radii is dominated by the geometrical effect, and outflows further reduce the net inflow.
}
\label{fig3}
\end{figure}

Figure~\ref{fig3} shows the net gas inflow rate across spherical shells about 
the stellar center of mass. Local inflow rates exceeding
10~M$_{\odot}$~yr$^{-1}$ are seen beyond the inner 1.5~kpc, dominated by clump
migration. The inner kpc is dominated by smooth inflow at a rate of $\sim 5\sy$.
The mild decline between $1\kpc$ and $50\pc$ can be partly explained by a
geometry effect, as the sphere radius becomes smaller than the disk thickness.
This is demonstrated by the inflow rate through cylindrical boundaries 
being rather constant with radius (Fig.~3, bottom panel). A
large fraction of the inflow takes place above and below the mid-plane in the
thick gaseous disk. An inflow rate of $1-2\sy$ nevertheless persists through the
central few parsecs across spherical boundaries. We notice that non-negligible
outflows, presumably from stellar feedback, further reduce the net inflow rate
in the central tens of pc. A comparison of runs HR1 and MR1 indicate no
resolution effects on scales of 10-100~pc.

This gas inflow fuels star formation in the bulge. Galaxy HR1 at $t=300$~Myr has
a total SFR of 132~M$_{\sun}$~yr$^{-1}$, of which 16~M$_{\sun}$~yr$^{-1}$ is in
the central kpc. This is less than the SFR of 42 and 31~M$_{\sun}$~yr$^{-1}$ in
clumps A and B, but comparable to the 14~M$_{\sun}$~yr$^{-1}$ in clump~C. 
At $t$=450~Myr, the stellar bulge mass has
grown from $9.8\times 10^{9}\,\msun$ to $2.2\times 10^{10}\,\msun$.
The persistence of star formation in the bulge as a result of disk instability is
consistent with observations showing that bulges in clumpy disks are bluer than
bulges in smooth disks \citep{E09}.

\newpage

\section{AGN obscuration in gas-rich disks}
\label{sec:obscuration}

The central BH resides in a dense and thick gaseous disk (Fig.~\ref{fig1}), which could obscure an AGN. To quantify this, we computed the column densities over 60 random lines of sight through the centers of galaxy HR1 at $t$=330~Myr and HR2 at $t$=380~Myr. The results are insensitive to the choice of center, selected to be either the center of mass of old stars or the point of maximum vorticity. The foreground hydrogen column density $N_\mathrm{H}$ was estimated over a cross-section corresponding to the minimum Jeans length, $4 \times \epsilon_\mathrm{AMR}$.

\medskip

\begin{figure*}
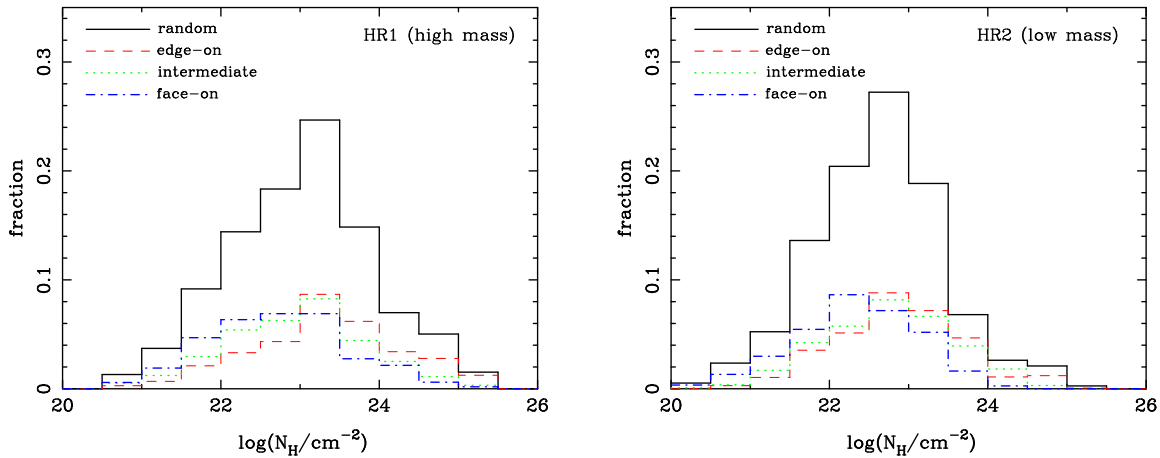

\centering
\includegraphics[width=6cm,angle=270]{nh1new.eps}\hspace{.8cm}
\includegraphics[width=6cm,angle=270]{nh2new.eps}
\caption{Distribution of gas column density along random lines of sight in front of a central AGN. We show separately the distributions for face-on ($i<30^\circ$), intermediate ($30^\circ<i<60^\circ$) and edge-on ($i>60^\circ$) orientations, the total fraction in each of these sub-samples being normalized to 33\% for clarity.} \label{fig4}
\end{figure*}

Figure~\ref{fig4} shows the distribution of $N_\mathrm{H}$ for runs HR1 and HR2 and for various sets of inclinations. 15\% of the lines of sight reach Compton thickness, $\log(N_\mathrm{H}) \geq 24.1$, and 50\% have $\log(N_\mathrm{H}) \geq 23$, i.e., strong obscuration. The obscuration tends to be higher in the more massive disk HR1, where Compton thickness can be reached even for face-on orientations owing to the $\sim 1$~kpc thickness of the gas disk. The lower-mass galaxy HR2, perhaps more representative at $z<2$, can reach Compton thickness almost exclusively for edge-on orientations. 

If most gas in the central parsec lies in a torus, unresolved in our models, the obscuration on edge-on projections may be further enhanced and the dependence of obscuration on inclination may be stronger. Significant obscuration would still be present along most lines of sight, because the column densities measured in the simulations are on average {\em not} dominated by the central 10~pc.

\section{Discussion: Accretion onto the black hole}

\def\Mbar{M_{\rm bar}}
\def\MBH{M_{\rm BH}}
\def\Mblg{M_{\rm blg}}
\def\Md{M_{\rm d}}
\def\MdotBH{\dot{M}_{\rm BH}}
\def\Mdotblg{\dot{M}_{\rm blg}}
\def\Mdotd{\dot{M}_{\rm d}}
\def\MdotE{\dot{M}_{\rm E}}

The isolated simulations presented here, using high gas fraction representative of $z\sim 2$ disks, have a disk inflow rate consistent with cosmological simulations of high-redshift stream-fed galaxies \citep{ceverino}. They reach resolutions better than 2~pc, hence resolving the nuclear inflow down to scales at which other processes drive the actual small-scale BH accretion \citep{combes00}. Only a fraction of the inflowing gas needs to make it all the way into the BH. The issues are similar to the case of merger-driven fueling, namely getting rid of angular momentum while avoiding excessive star formation and AGN feedback. This is beyond the scope of our paper, and we limit ourselves to heuristic estimates.

One can assume that the local relation between the BH mass $\MBH$ and the bulge 
properties (mass $\Mblg$ and velocity dispersion $\sigma$) is crudely valid at high redshift. If we adopt $\MBH/\Mblg
\sim 10^{-3}$ at $z\!=\!0$, assume it scales as $\sigma^2$ and allow a cosmological scaling $\sigma^2 \propto (1+z)$, then we obtain $\MBH/\Mblg \sim 3 \times 10^{-3}$ at $z\! \sim\! 2$. A $z\! \sim \! 2$ galaxy of baryonic mass
$10^{11}\,\msun$ hosts a BH of $\sim \! 10^8\, \msun$, while in a gravitationally unstable steady state it is typically half bulge and half disk (DSC09). According to \equ{Mdot}, the average inflow rate through the disk into the bulge is $\Mdotblg \sim 12.5 \sy$. If the ratio of average accretion rates into bulge and BH follow the ratio of the corresponding masses, the BH accretes on average at $\MdotBH \sim 0.04 \sy$, which is 1-2\% of the Eddington rate (assuming a $0.1$ efficiency for mass-to-energy conversion). The corresponding bolometric luminosity is $2 \times 10^{44}\ergs$. With typically 1--5\% in X-rays, we estimate on average $L_{\rm X} \sim 10^{42-43}\ergs$, scaling with galaxy mass and with $(1+z)^{2.5}$.
While the average luminosity would be modest, short episodes of higher accretion rate, possibly up to the Eddington level, occur during the central coalescence of migrating giant clumps -- which could also bring with them seed BHs \citep{EBE08}.

\medskip

At very high redshift, the same process could feed brighter AGN in rare massive systems \citep{DM11}, provided that unstable disks indeed form there. The average inflow rate increases with redshift (eq.~2), such that a $z\!=\!6-10$ disk can support continuous accretion at the Eddington rate in self-regulated quasar mode if $\MBH/\Md$ is $\sim\!0.04-0.08$, for the same $\sim\!10^{-3}$ fraction of the inflow rate accreting onto the BH, and assuming outflows would not affect the cold gas streaming inward. 
Then a seed BH of $10^5\,\msun$ at $z \!\sim \! 10$ can grow exponentially by 10 e-folds to $2\times 10^9\,\msun$ at $z\! \sim\! 6$, possibly explaining very massive BHs growing in gas-rich systems at $z\!>\!5$ \citep{mortlock,treister11}. 
The typical baryonic mass of the host galaxy is $\sim \! 10^{11}\,\msun$, and if the halo mass is a few times $10^{12}\,\msun$, the comoving number density of such bright $z>5$ AGNs would be $\sim \!10^{-8}$ to $10^{-6}\,\Mpc^{-3}$. 

\section{Conclusion}

Violent gravitational instability in high-redshift disk galaxies naturally leads
to the growth of a bulge and a central BH; it can both feed an AGN and obscure it. 
This stems from the developing
picture where many star-forming galaxies at high redshift are 
rotating disks with high gas fractions continuously fed by cold streams. 
Such disks develop instability that involves transient perturbations and giant 
clumps. Gravitational torquing among these perturbations lead to mass inflow, which 
provides energy for maintaining the strong turbulence required for  
self-regulated instability at $Q\! \sim \! 1$.
The cosmological inflow rate along cold streams sets an upper limit, but the actual inflow rate toward the nucleus is determined by the disk instability on galactic scales.

High-redshift disks are very different from nearby spirals, the key difference 
being the much higher gas fraction. The instability then operates on short 
dynamical timescales, and drives an intense inflow through the disk at the 
level of $\sim \!10\sy M_{\rm b,11} (1+z)_3^{3/2}$.
In low-redshift disks, the instability is a secular process limited to 
non-axisymmetric modes with considerably weaker torquing: 
bars are invoked in some AGN models \citep{fanidakis, begelman} but need
about ten rotation periods to convey some gas inward \citep{BCS05}.
Gravitational torquing in high-$z$ disks is much more efficient and involves richer gas reservoirs.
  
\medskip

Disk instability at $z \! \sim \! 2$ can thus funnel half of the disk gas toward the center in 2~Gyr. This is similar to the mass inflow in a major merger \citep{H06}, but spread over a ten times longer period, resulting in a lower average AGN luminosity, with higher duty cycle, and high obscuration. 
This process could dominate, as the gas mass in the relatively smooth cosmological streams is larger than that associated with mergers -- less than 10\% of the cosmological infall is in major mergers \citep{dekel09, lhuillier11}. The main prediction is thus that many high-$z$ AGNs should be hosted by star-forming disk galaxies, composed of clumpy disks and growing spheroids. 
Many would be moderately luminous, detectable by {\it Chandra} deep surveys at $z\!=\!2$ if un-obscured, but the high expected obscuration makes their detectability dubious, especially for galaxies below $10^{11}\,\msun$. These BHs can grow inside classical bulges \citep{EBE08b}.

Observationally, the fraction of X-ray-detectable AGN indeed declines with decreasing stellar mass \citep{xue10}. Many AGN are highly absorbed, individually undetectable in X-rays \citep[e.g.,][]{daddi07}, and many of these obscured AGN have moderate intrinsic luminosities and lie in star-forming galaxies \citep{J11}. Host galaxies of moderate-luminosity AGNs do not show an excess of major-merger signatures \citep{grogin05,gabor09} and often have disky morphologies \citep{schawinski11}. Small mergers may also help feeding the BH while leaving the disk intact \citep{shankar10}, but they are a component of the cosmological streams, and can be considered to be part of the inflow within the disk. 

\medskip

The violent instability phase should end when the cosmological accretion rate declines and the system becomes stellar dominated. We expect less massive galaxies to remain unstable for longer times, because they retain higher gas fractions --- one of the aspects of the downsizing of star formation \citep[e.g.,][]{J05}. This can result from the regulation of gas consumption in smaller galaxies \citep{DekelSilk86, KD11} and from the continuation of cold accretion to lower redshift for lower-mass halos \citep{DB06}. It induces an inverse gradient of gas fraction with galaxy mass (as observed, \citealt{kannappan04}) and a downsizing in gravitational instability which could result in downsizing of BH growth. This is a longer growth phase into later redshifts in lower-mass galaxies, with $\sim \! 10^{11}\msun$ galaxies growing BHs mostly at $z > 1$ in their unstable disk phase, while clumpy disks of $\sim \! 10^{10}\msun$ may show AGN activity even after $z \! \sim \! 1$.

\acknowledgments

Simulations were performed at CCRT and TGCC under GENCI allocation 2011-GEN2192. We acknowledge discussions with Fran\c{c}oise Combes, Bruce Elmegreen, Tiziana Di~Matteo, James Mullaney, a constructive referee report, and support from grants ERC-StG-257720, CosmoComp ITN, ISF 6/08, GIF G-1052-104.7/2009, NSF AST-1010033 and a DIP grant.

{}

\end{document}